\documentclass[superscriptaddress,aps,prl,twocolumn,preprintnumbers,amsmath,amssymb,floatfix,groupedaddress,10pt]{revtex4-1}
\usepackage[dvips]{graphicx}
\usepackage{dcolumn}
\usepackage{bm}
\usepackage{xr}
\usepackage{color}
\usepackage{amsmath}
\usepackage{hyperref}
\usepackage{floatrow}
\usepackage[caption=false]{subfig}
\newcommand{\ket}[1]{\ensuremath{\left|#1\right\rangle}}
\newcommand{\oa}{\hat{a}}
\newcommand{\ob}{\hat{b}}
\newcommand{\oA}{\hat{A}}
\newcommand{\oB}{\hat{B}}

\begin{document}
	
\title{Generating strong anti-bunching by interfering with coherent states}
\author{Rajiv Boddeda, Quentin Glorieux, Alberto Bramati and Simon Pigeon}
\affiliation{Laboratoire Kastler Brossel, Sorbonne Universit\'{e}, CNRS, ENS-PSL Research University, Coll\`{e}ge de France, 4 place Jussieu, 75252 Paris, France}

\begin{abstract}
	The second order correlation function is traditionally used to characterize the photon statistics and to distinguish between classical and quantum states of light. In this article we study a simple setup offering the possibility to generate strong anti-bunched light. This is achieved by mixing on a beam splitter a coherent state with a state with a non-negative Wigner function, such as squeezed states or weak Schr\"{o}dinger cat states. We elucidate the interference mechanism generating such strong antibunching and relate it to unconventionnal photon blockade. We also detail how this effect can be used to measure weak squeezing.
\end{abstract}

\maketitle
\section{Introduction}
Non-classical photon sources play a crucial role in emerging quantum technologies. Given the robustness of quantum states of light, photons are the best candidates for applications in the field of quantum communication and quantum cryptography. \cite{OBrien2009,Gisin2007}.
One common way of characterizing the non-classical nature of light sources is by measuring the second order correlation function of the field intensity, $g^{(2)}(\tau)$.
The value of this function at $\tau$ = 0 for classical light is larger than one and is equal to one for a coherent state. Therefore $g^{(2)}(0) < 1$ is seen as a clear signature of the non-classical nature \cite{Paul1982}. It reveals that the emission temporal statistics of photons is sub-poissonian (more ordered in time than a coherent source) such as we speak about anti-bunched emission.

Anti-bunched states of light also have applications beyond quantum technologies in fields such as super-resolution microscopy. It is possible to go beyond the diffraction limit by taking advantage of ordered photon emission such as strongly antibunched light \cite{Schwartz2012,Schwartz2013}.

Such states of light are typically obtained using nanoemitters such as semiconductor quantum dots \cite{michler2000quantum,pisanello2010room} or nitrogen vacancy centers in diamond \cite{Kurtsiefer2000}, and allows, for the most efficient platform, for single photon sources \cite{aharonovich2016solid}.
The anti-bunched nature of the light emitted by these devices derives directly from the extreme confinement of matter excitations. Another approach is to modify a light beam in order to obtain a photonic state with a sub-poissonian statistics i.e. $g^{(2)}(0) < 1$. It typically requires a strongly nonlinear medium with enough effective photon-photon interactions to observe the photon blockade \cite{Deutsch1997,Birnbaum2005}.
However, it has been recently proposed that this statistics can also be achieved, by combining weak photon-photon interactions and optical path interference \cite{SavonaPRL,SavonaReview,Bamba2011}. This configuration, known as unconventional photon blockade, has been realized experimentally and validated in a quantum dot system \cite{Snijders2018} and in a superconducting circuit \cite{Vaneph2018} paving the way to efficient source of anti-bunched light.

In this manuscript, along this line, we emphasize that any kind of weakly squeezed states can be used to generate strong anti-bunching. This observation, pointed out already four decades ago \cite{Paul1982,Ritze1979}, is revised and analyzed to show that one can mix a pure state with correlation function $g^{(2)}(0) \ge 1$ with a coherent state ($g^{(2)}(0) = 1$), to create anti-bunched states with $g^{(2)}(0) \simeq 0$. A beamsplitter or a cavity in reflection are capable of generating these non-classical states. We elucidate how interferences at the level of Fock state component of considered states of light can lead to such strong anti-bunching. It is directly connected to unconventional photon blockade \cite{SavonaPRL}, in contrast our proposal does not resort to a sophisticated time filtering techniques.

\begin{figure}[t]
	\centering
	\includegraphics[width=0.8\linewidth]{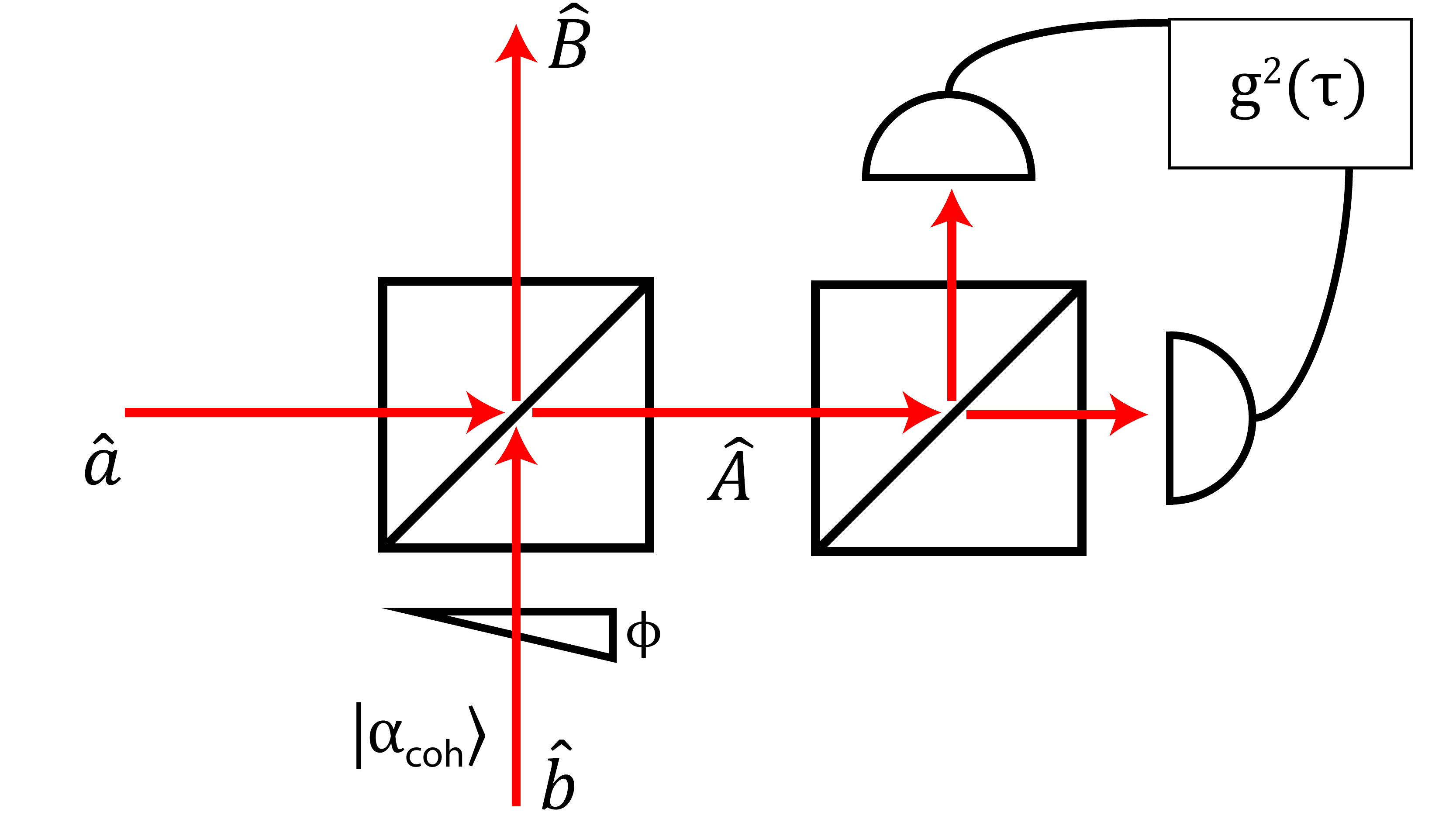}
	\caption{Schematic setup for the creation of anti-bunched states.
	The two input operators are indicated by $\oa$ and $\ob$, and the corresponding output operators are denoted by $\oA$ and $\oB$. The input mode $\ob$ consists of a coherent state with a tunable phase $\phi$ relative to $\oa$. Several input states $\oA$ are studied in this paper. The correlation function is computed on the output mode $\oA$ in the same way as it is done experimentally i.e. using a beamsplitter and two photo-detectors as represented.} \label{fig:BS}
\end{figure}

However, using the input-output relation \cite{Walls2008} and the factorization property \cite{Grankin2015PRA} in the cavity reflection setup case, one can show that $g^{(2)}(0)$ cannot go below $0.5$, for a cavity filled with a weak Kerr non-linear media. This limit can be overcome using a different system \cite{Grankin2015PRA}, but require far more complex setup than the one proposed here presented in Fig. \ref{fig:BS}.  


The setup is based on two input beams mixed on a beamsplitter. Relative phase between both inputs can be tuned via a delay line and we focus on the statistics of one of the two input using standard coincidence measurement scheme \cite{hong1987measurement}.  
In what follows, we first define the model used and then explore the $g^{(2)}(0)$ obtained when mixing a coherent state with a perturbed coherent state. We consider different phase modified coherent states and also squeezed and Schr\"{o}dinger cat states of light. In the latter case, we derive the exact result and propose an application. Finally, we discuss the advantage of the present setup over the other unconventional photon blockade setups.  

\section{Model and methods}

The statistics of a stationary field in a given electromagnetic mode can be quantified using the second order correlation function \cite{loudon1973quantum} :
\begin{equation} \label{eq:g2}
    g^{(2)}(\tau) = \frac{\langle\oA^\dagger(0)\oA^\dagger(\tau)\oA(\tau)\oA(0)\rangle}{\langle\oA^\dagger(0)\oA(0)\rangle^2}\,,
\end{equation}
where $\oA$ is the annihilation operator in the mode in which we would like to measure $g^{(2)}(\tau)$.

We consider a pure quantum state of light in a given quantized mode, which can be expressed in the Fock state basis using $\ket{\psi} = c_0 \ket{0}+c_1\ket{1}+ c_2\ket{2}+ c_3\ket{3}...$, \cite{scully_zubairy_1997} where the coefficients, $c_i$, can be time dependent. 
The second order correlation function of this state can then be calculated using :
\begin{equation} \label{eq:g2approx}
g^{(2)}(\tau) = \frac{2|c_2|^2+6|c_3|^2+..}{(|c_1|^2+2|c_2|^2+3|c_3|^2...)^2}\,.
\end{equation}
A sub-poissonian statistics will imply that numerator of Eq. (\ref{eq:g2approx}) is smaller than its denominator. For a weak amplitude state where $|c_n|^2$ vanishes quickly with $n$ and anti-bunching can be achieved either by increasing $|c_1|^2$ or minimizing $|c_2|^2$. Notice that, both single and two photon component are in general strongly related. This proposal proposes to specifically cancel $|c_2|^2$ using interference effect. In what follows, if any truncation is applied to Eq. \ref{eq:g2approx}, the convergence with respect to the truncation is always verified.

\subsection{Input-output relations of a beamsplitter}
We consider a simple setup as shown in Fig. \ref{fig:BS}, in which we mix a coherent state in mode $\hat{b}$ with another state in mode $\hat{a}$ on a beamsplitter. We denote the two output modes of the beamsplitter as $\oA$ and $\oB$ and we evaluate the value of $g^{(2)}(0)$ in the mode $\oA$. This setup is similar to the one studied in Ref. \cite{Ritze1979}.
The states considered in the mode $\hat{a}$ will be of different forms, but always relatively close to a coherent state. Typically no state with negative Wigner function will be considered.
We consider a beamsplitter with no losses and use the input-output relations to evaluate the output state. Moreover, we have the possibility of an additional phase shift of $\phi$ on the input arm $\ob$, to explore the interference effect. 

The two outputs of the beamsplitter can be written as 
\begin{eqnarray}
\oA &=& \sqrt{T}\oa+\sqrt{R}\ob e^{i\phi \pi} \nonumber \\
\oB &=& -\sqrt{R}\oa+\sqrt{T}\ob e^{i\phi \pi},\label{eq:BS}
\end{eqnarray}
where, $R$ and $T=1-R$ are the reflectance and the transmittance of the beamsplitter respectively and $\phi\in[0,2]$ (note that $\phi$ is normalized to $\pi$).

\section*{Phase modified coherent states}

In this section, we consider phase modified coherent states, i.e.  normal coherent states  to which we apply different dephasing to different Fock number states in the Fock state basis.
By considering such an input state in mode $\oa$ plus a coherent state in mode $\ob$, we observe clear and strong anti-bunching in $\oA$ for certain parameters.

\begin{figure}[t]
  \centering
  \includegraphics[width=\linewidth]{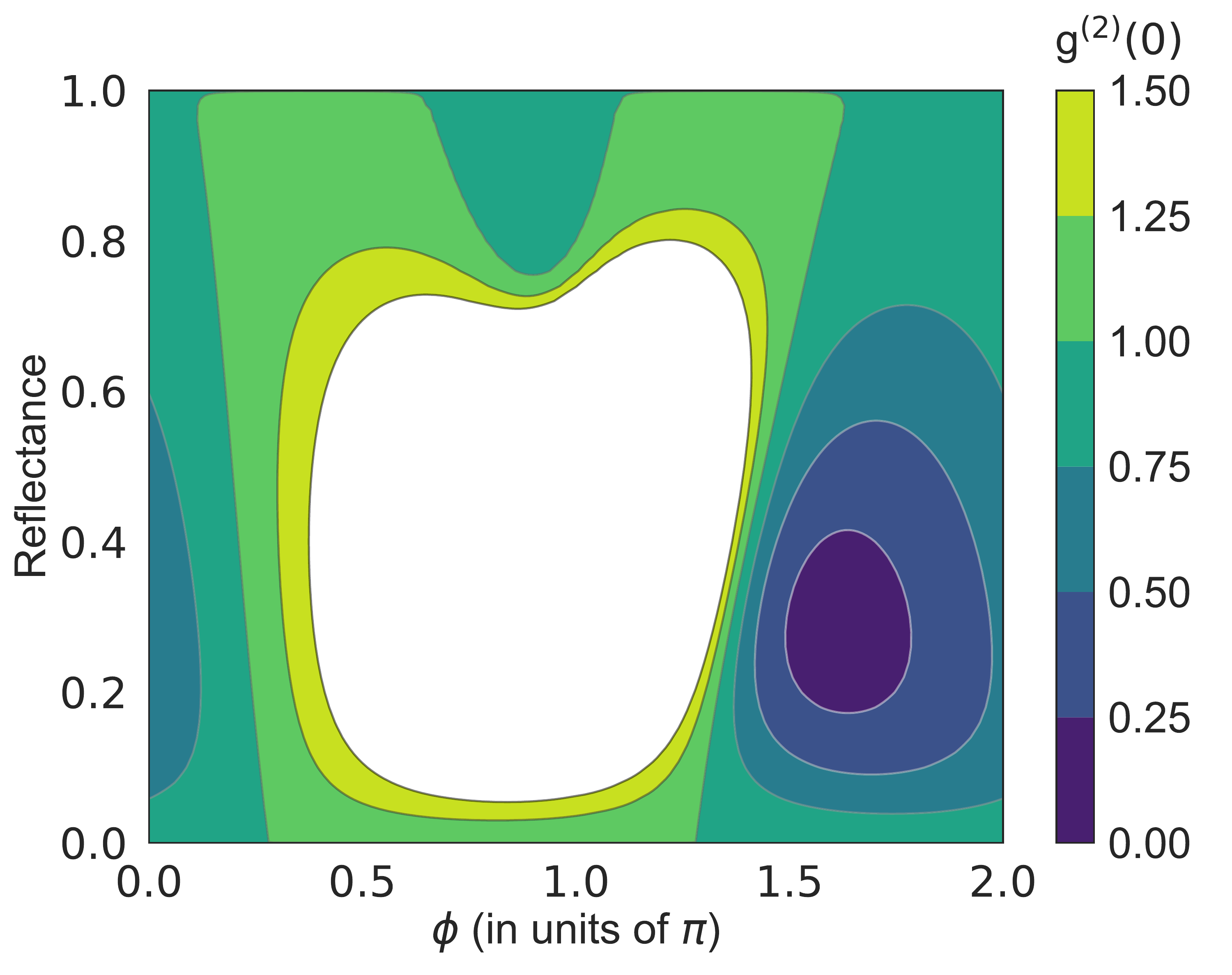}
    \caption{The correlation function of the output state as a function of the reflectivity and $\phi$, for a given amplitude of $\alpha = 0.3$. One can observe how the correlation function goes below 0.5 for low reflectivities and around $\phi = 1.6$. The white region in the plot corresponds to $g^{(2)}(0) >$ 1.5.}
  \label{fig:VacuumModified}
\end{figure}

Let's consider, for pedagogical purposes, a coherent state with the two photon Fock component dephased with respect to the other Fock components and interfering with a coherent state. This state  can be written in the following form
\begin{equation}\label{eq:modcohstate}
 \ket{\psi} = \sum_{n\ge 0} \frac{\alpha^n}{\sqrt{n!}}e^{-\frac{|\alpha|^2}{2}}e^{i\frac{\pi}{2}\delta_{n,2}} \ket{n}\,,
\end{equation}
where $\delta_{n,2}$ is the Kronecker delta function and $\alpha$ is the amplitude of the state. 

\begin{figure*}[ht]
  \sidesubfloat[\label{sfig:onenonlinear}]{%
  \includegraphics[width =0.45 \linewidth]{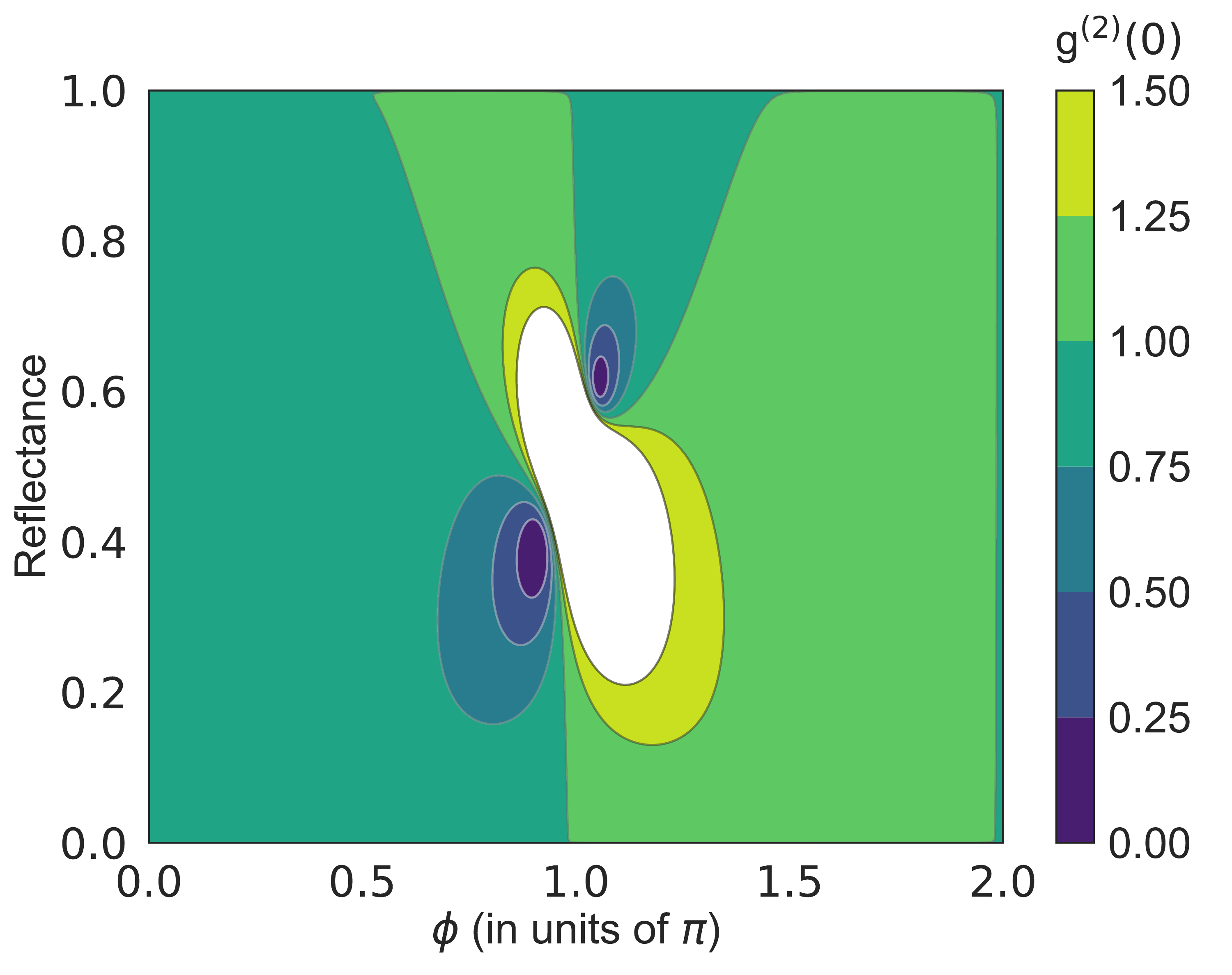}%
}\hfill
  \sidesubfloat[\label{sfig:photonnonlinear}]{%
  \includegraphics[width = 0.45\linewidth]{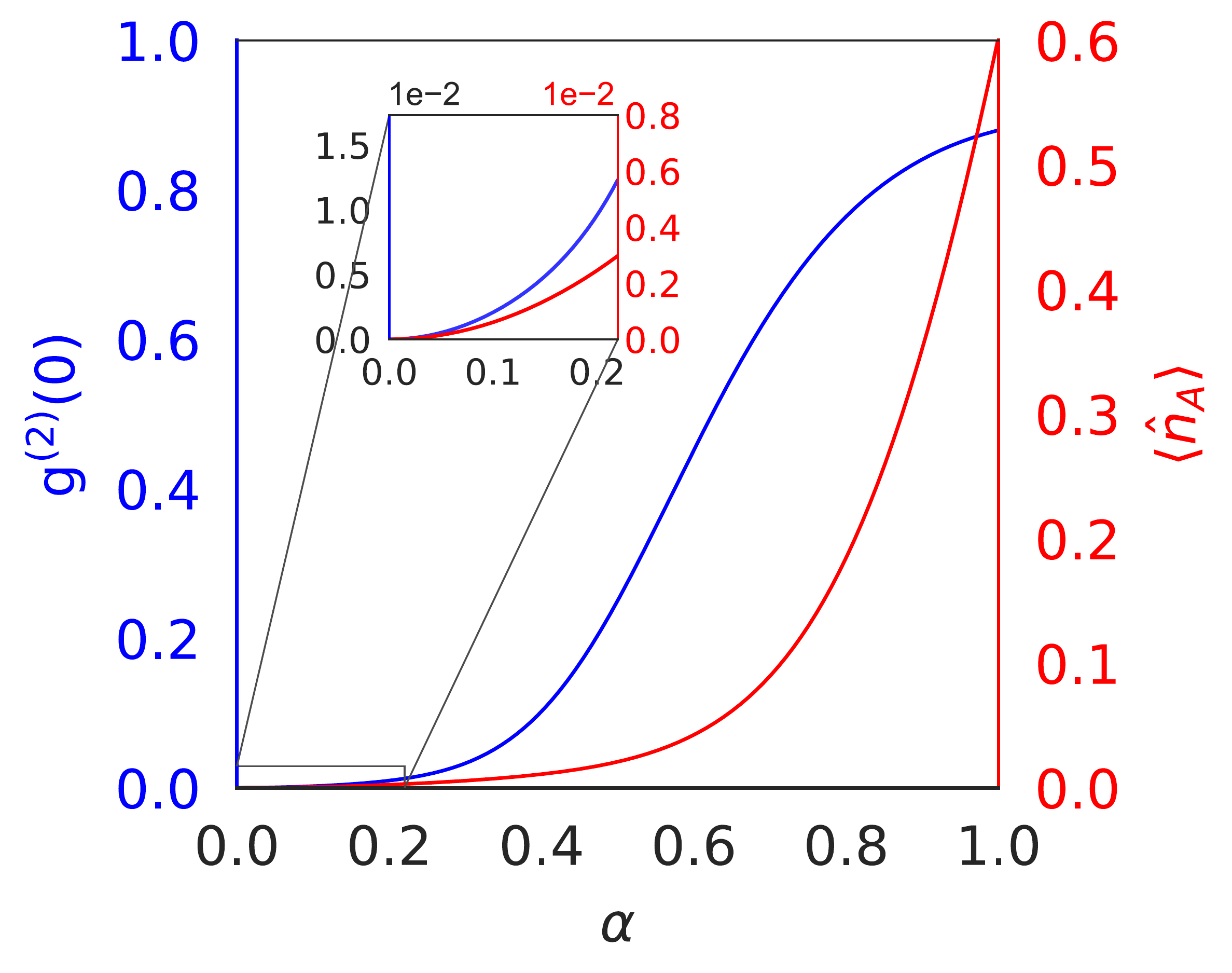}%
}
\caption{The correlation function, $g^{(2)}(0)$, and photon number, $\langle \hat n_{\text{A}}\rangle$, as a function of the reflectivity and the phase angle $\phi$ when considering one coherent state modified through non-linear medium as given in Eq. (\ref{eq:nonlinear}) in the input mode $\hat{a}$, and another coherent state with same amplitude $\alpha = 0.3$ in the input mode $\hat{b}$. 
We assume that the non-linear coherent state $\ket{\alpha}$ has propagated through a weak non-linear medium with the parameter $\chi t = 0.05$.
(a) Correlation function at zero delay for $\hat{A}$.
(b) The minima of the $g^{(2)}(0)$ and the corresponding average photon number in mode $\hat{A}$, as a function of the coherent state amplitude in the input mode for optimal $\phi$ and $R$. Inset in (b): magnification of the two plots.}
\label{fig:OutputInterference}
\end{figure*}

In Fig. \ref{fig:VacuumModified} the $g^{(2)}(0)$ function is plotted  as a function of the relative dephasing $\phi$ and the reflection coefficient R. One can observe that mixing this phase modified state, as defined in Eq. (\ref{eq:modcohstate}), with a coherent state allows us to generate perfect anti-bunched states in the output mode $\hat{A}$.
For example, by considering a beamsplitter with 15\% reflectivity and a relative phase of approximately $\pi$, for an amplitude of the state less than 1, pure anti-bunched state can be generated. 
However we found that the minimum of the correlation function increases exponentially as a function of the number of photons in the input coherent state, $|\alpha|^2$. 
This example clearly shows the strong dependence of the output correlation function on the relative phase between the different Fock components of the incoming states. \\

In practice, states such as defined in Eq. (\ref{eq:modcohstate}) are unphysical. In the following, we consider another type of state that can be realized experimentally, such as the one obtained after letting a coherent state propagate through a purely $\chi^{(3)}$ non-linear medium. 
This state can be written in the following form \cite{Banerji2001}
\begin{align}\label{eq:nonlinear}
    \ket{\alpha(t)} &= \sum_n \exp{\left(-i \hat{H}_{nl}t\right)}c_n\ket{n} \nonumber\\
    &= \sum_n \exp{\left(-i \chi^{(3)} t (n^2-n)\right)} c_n\ket{n}\,,
\end{align}
with $\hat{H}_{nl}$ the non-linear Hamiltonian and $\chi^{(3)}$ the third order non-linear susceptibility of the medium.

One can observe that in this case, non-linearity implies that the phase of each Fock component evolves at a different rate, accumulating a different phase while propagating.
Interfering such a state with a coherent state on a beamsplitter with an adequate relative phase between both beams and a proper reflection coefficient, leads to strong anti-bunching as shown in Fig. \ref{fig:OutputInterference}. 

We show in the panel (a) of Fig. \ref{fig:OutputInterference} the numerical evaluation of the corresponding second order correlation of the output field $\oA$, as a function of the relative phase $\phi$ between the two input beams and the beamsplitter reflection coefficient $R$. Even though the antibunched region is smaller than in the previous example, we can clearly observe two dips in the correlation function map. 
Here, for simplicity, we consider both input sate with the same amplitude  $\alpha = 0.3$. We took $\chi^{(3)} t = 0.05$ for the non-linearity required to produce the non-coherent state, which corresponds to relatively small non-linearity easily achievable experimentally \cite{Hickman2015}. 
One would observe the same effect in the other output mode ($\oB$) but for a symmetric set of parameters ($R'= 1-R$ and $\phi'= 1- \phi$).
An important thing to note is that the strong anti-bunching observed corresponds to a non-vanishing output intensity.
Indeed, the prospect to find the smallest $g^{(2)}(0)$ is relevant only if the output state is not vacuum. 
In the present case, we find that $g^{(2)}(0)=0.03$ while $\langle \hat{n}\rangle \approx 0.006$ for $\alpha=0.3$.
To that purpose we show in the panel (b) of Fig. \ref{fig:OutputInterference} how the second order correlation and the photon number in the output arm vary with the amplitude $\alpha$ at the optimal condition. Here we assume that both the input states have the same amplitude. We clearly see that $g^{(2)}$ increases when increasing the input field amplitude $\alpha$. It increases faster than the mean number of photon of the output field $\langle \hat{n}_A \rangle$. However, until $\alpha=0.5$ we have strong anti-bunching ($g^{(2)}< 0.5$) and for $\alpha<0.2$ we even found very strong anti-bunching ($g^{(2)}<0.01$) with non-zero output amplitude (up to $\langle \hat{n}_A \rangle \approx 0.003$) (inset of the panel (b) of Fig. \ref{fig:OutputInterference}). 

The setup described can be realized experimentally using a Mach-Zehnder interferometer with a non-linear medium in one of the arms.
The phase $\phi$ is modified by tuning one arm of the interferometer.\\

Other non-classical states can be used along the same scheme. In the following we will focus on states only composed of even Fock components.

\begin{figure}[t]
	\centering
	\includegraphics[width=0.9\linewidth]{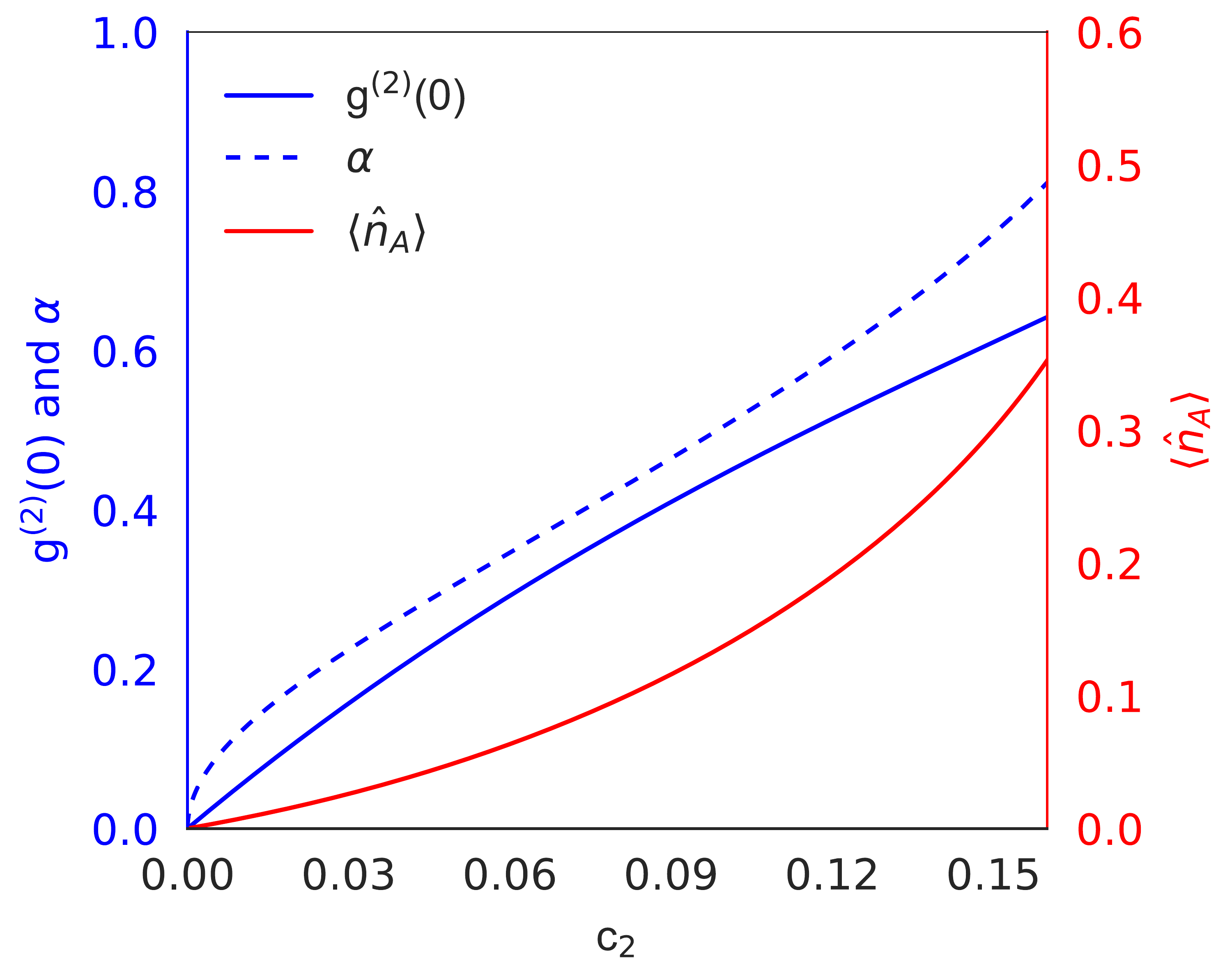}
	\caption{Second order correlation function and photon number at the output arm, as a function of the amplitude of the two-photon coefficient of the superposition state, $c_2$. The dashed blue curve corresponds to the amplitude of the coherent state to obtain the minima of $g^{(2)}(0)$ at the output arm of a 50:50 beamsplitter, the solid blue curve corresponds to $g^{(2)}(0)$, and the red curve represents the average photon number in that arm.} \label{fig:twophoton}
\end{figure}

\section*{Even Fock states}
In the above section, we have shown that when we modify the relative phase between the Fock components of a coherent state, one can produce strongly anti-bunched states. 
In this section, an alternative route is explored to demonstrate that this phenomenon can also be observed when the amplitudes of Fock components are modified.
We consider examples in which the odd photon components are reduced compared to the coherent states, similar to the case of Schr\"{o}dinger even cat states. 
In these cases, mixing even a slightly squeezed state with a coherent state will lead to the observation of strongly anti-bunched statistics.
Despite the difference in the approach proposed here with respect to the previous cases, the underlying idea is identical: interfering the two photons Fock component of the squeezed states with that of the coherent state, can generate strong anti-bunching.

\subsection*{Superposition of a vacuum and two photon states}
We begin with a simple example: a normalized weak two-photon pure state in superposition with vacuum defined by $\ket{\psi} = c_0\ket{0}+c_2\ket{2}$. In this case we have $g^{(2)}(0)=1/(2|c_2|^2)\ge 0.5$. 
If we consider this state in one of the input modes of the beamsplitter ($\oa$) and a coherent state in the other input mode, i.e. both the input states with $g^{(2)}(0)\ge0.5$, then the output state in the mode $\hat{A}$ can exhibit strong sub-poissonian statistics such as $g^{(2)}(0)\ll 0.5$.
In Fig. \ref{fig:twophoton}, we show how the correlation function (solid blue line) changes as a function of the amplitude of the two photon state $c_2$ for a 50:50 beamsplitter. The $g^{(2)}(0)$ is minimized for each value of $c_2$ by tuning $\alpha$, the amplitude of the input coherent state in the mode $\hat{b}$. Corresponding $\alpha$ is represented in Fig. \ref{fig:twophoton} in dashed blue line. The red line represents the mean photon number in the output mode $\oA$. 
One can clearly observe that the correlation function $g^{(2)}(0)\le0.5$ for $c_2\le 0.1$, even though the input superposition state in $\oa$ has $g^{(2)}(0)>50$ for $c_2 < 0.1$.

This example is counter intuitive because it shows that mixing on a beamsplitter a superposition state with only vacuum and two-photon component may lead to a strong sub-poissonian statistics which is a  signature of a single photon states.


\subsection*{Schr\"{o}dinger cat states}
This formalism can be extended to Schr\"{o}dinger cat states. 
Schr\"{o}dinger cats are formed by the superposition of two coherent states with opposite phase. Depending on whether the coherent states are added or subtracted, the resulting state is either referred to as even or odd cat state, respectively. 
In the weak amplitude case, if one evaluates the correlation function of odd cat states using the Eq. (\ref{eq:g2}), one can find that they exhibit anti-bunching, since they only contain odd Fock states. Using such states in the present scheme and mixing them with a coherent state will always lead to increase of $g^{(2)}(0)$ in the output arm. As the odd cat state does not encompass two-photon component, mixing it with a coherent state always leads to an output state with residual two photons components and hence a higher $g^{(2))}(0)$ in the output than in the input. To reduce $g^{(2))}(0)$, more complex schemes can be implemented to keep the $c_2$ at the output close to zero and reduce the three photon component at the same time.   
However, this is different with the even cat states.
The even cat states can be written as 
\begin{equation}
    \ket{\text{cat}_e} \propto \ket{0} + \frac{\alpha_{sch}^2}{\sqrt{2}}\ket{2} + \frac{\alpha_{sch}^4}{2}\ket{4}+...\,,
\end{equation} 
where $\alpha_{sch}$ is the cat state amplitude. 
Since the expressions are not tractable analytically, we evaluate the output correlations numerically using the QuTiP toolbox \cite{Johansson2013}.

\begin{figure}[t]
	\centering
	\includegraphics[width=\linewidth]{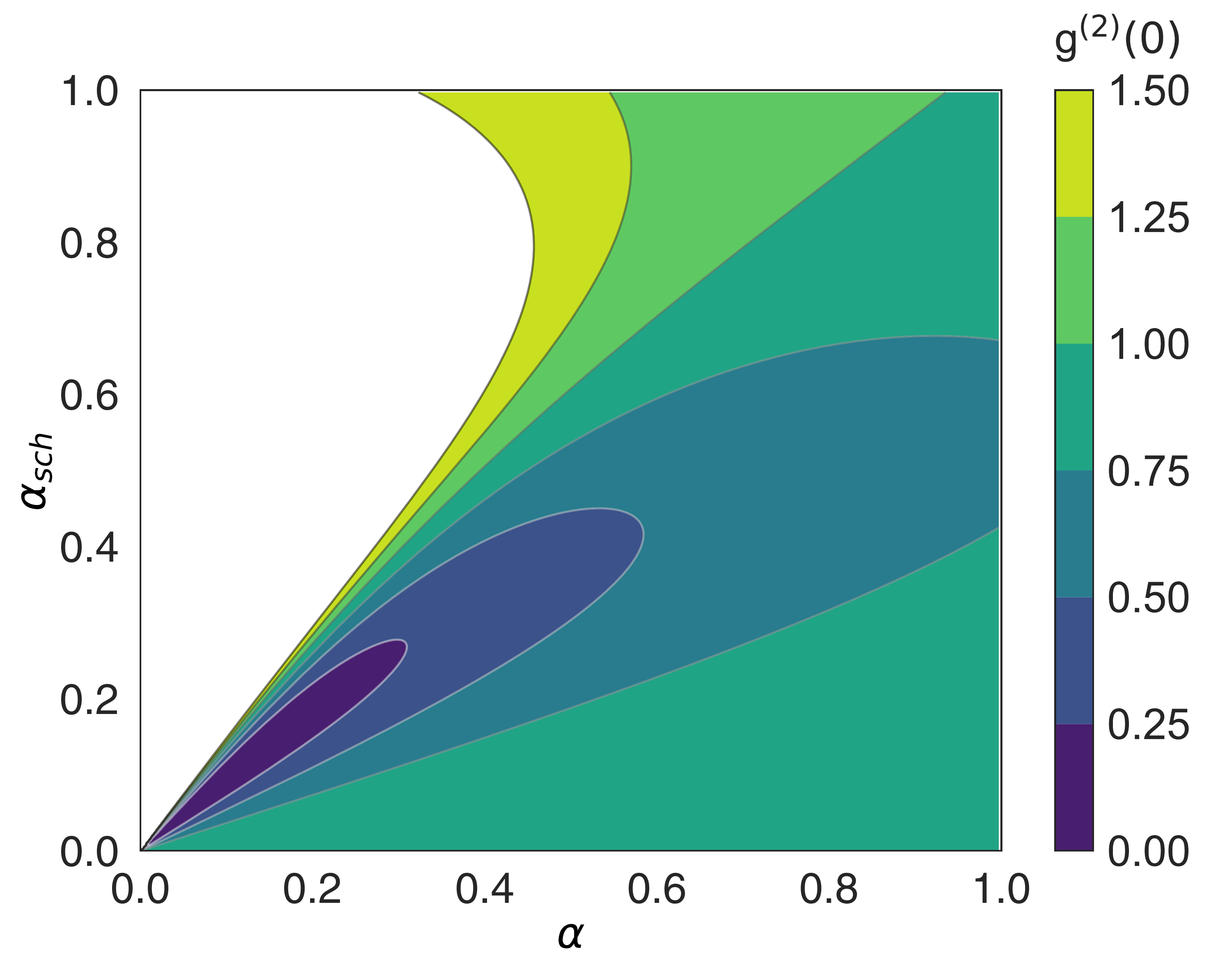}
	\caption{Contour plot of the second order correlation function at the output arm as a function of the amplitudes of the coherent state ($\alpha_b$) and the Schr\"{o}dinger cat state ($\alpha_{sch}$) in the two input arms. We consider a 50:50 beamsplitter.} \label{fig:SC_BothChanging}
\end{figure}

We assume the beamsplitter to be symmetric, i.e. 50:50. 
We represent in Fig \ref{fig:SC_BothChanging} how $g^{(2)}(0)$ varies as a function of $\alpha$, the amplitude of the coherent state in mode $\hat{b}$, and of $\alpha_{sch}$, the amplitude of the input cat state in mode $\hat{a}$. 
We see that if $\alpha_{sch} < \alpha $, it leads to bunching whereas  $\alpha_{sch} > \alpha $ leads to anti-bunching. 
Moreover, for $\alpha_{sch} \approx \sqrt{\alpha}/2$ we obtain the minimum of $g^{(2)}(0)$ where $g^{(2)}(0)\to 0$ with $\alpha \to 0$. Interestingly, the sensitivity of the correlation function (local derivative with respect to $\alpha_{sch}$ and $\alpha$) increases while $g^{(2)}(0)$ decreases for decreasing amplitudes.

There are several experimental techniques to generate Schr\"{o}dinger cat states \cite{Ourjoumtsev2006, Ourjoumtsev2007,Stobinska2007, Huang2015}. Characterizing such states is a very challenging task as it usually relies on full state tomography with high sensitivity. This simple scheme can offer an interesting alternative. A cat state of a given amplitude mixed with a coherent state can be uniquely accessed via the value of $g^{(2)}(0)$ of the output field. Moreover, variation of this value with respect to the amplitude of the input coherent field is also directly linked to the amplitude of the cat state. Both quantities, easily accessible, can be advantageously used to witness low amplitude Schr\"{o}dinger cat states.

It is known that at low amplitudes $\alpha_{sch}$, cat states converge to squeezed states which we now consider explicitly. 

\subsection*{Squeezed coherent states}
Squeezed coherent states have been observed in many different types of systems, such as parametric down conversion \cite{SqueezingPDC}, optical fibers \cite{Bergman91}, semiconductor lasers \cite{Machida1987}, four-wave-mixing \cite{glorieux2010double,glorieux2011quantum,corzo2013rotation}, etc.
In this part, we see how one can use squeezed coherent states to create anti-bunching.
If one considers a vacuum squeezed state, then it consists of only even Fock states \cite{Gong1990}.
Hence, squeezed coherent states should be similar to the case discussed previously. 

For a squeezed coherent state and a coherent state respectively in the input modes $\oa$ and $\ob$, the total input state can be written as
\begin{align}
|\psi\rangle_{\text{input}} = D(\alpha_a,\oa)S(\xi_a,\oa)|0\rangle_a \otimes D(\alpha_b,\ob)|0\rangle_b\,,
\end{align}
\noindent where $\oa$ and $\ob$ are the annihilation operators acting on the two input modes. 
The displacement and the squeezing parameters in the corresponding modes are denoted by $\alpha = |\alpha| e^{i\Phi}$ and $\xi = r e^{i \omega}$, with subscripts indicating the modes in which they act on, respectively. In the latter, $\omega$ denotes the squeezing angle.  
The squeezing operator is given by $S(\xi,\oa) = e^{\frac{1}{2}(\xi^* \oa^2 - \xi \oa^{\dagger 2})}$ and the displacement operator by $D(\alpha,\ob) = e^{(\alpha \ob^{\dagger} - \alpha^* \ob)}$ \cite{scully_zubairy_1997,Lvovsky2015}.

Using the beamsplitter relations in Eq. (\ref{eq:BS}), we can write the output state of the beamsplitter as 
\begin{align}
|\psi\rangle_{\text{out}}=D(\alpha_a,\sqrt{T}\oA - \sqrt{R}\oB) D(\alpha_b,e^{-i\phi}({\sqrt{R}\oA + \sqrt{T}\oB})) \nonumber \\
S(\xi_a,\sqrt{T}\oA - \sqrt{R}\oB)\left(|0\rangle_A \otimes |0\rangle_B \right)
\end{align}
Since the displacement operators always commute with each other, one can simplify the displacement part of the equation to $D(\alpha_b',\sqrt{R'}\oA + \sqrt{T}'\oB)$, in which 
\begin{align}
\sqrt{R'} &= \frac{\alpha_a}{\alpha_b'}\sqrt{T}+ \sqrt{R}\,, \\
\sqrt{T'}&= \sqrt{T}- \frac{\alpha_a}{\alpha_b'} \sqrt{R}\,, \\
\alpha_b' &= \alpha_b e^{i\phi}\,.
\end{align}
The squeezed state in the input mode is divided into two squeezed coherent states in the two output modes. The equation can then be simplified to 
\begin{align}\label{eq:sqoutput}
|\psi\rangle_{\text{out}} = D(\alpha_b,&\sqrt{R'}\oA + \sqrt{T'}\oB) \nonumber \\ &S(T \xi_a,\oA)S(R\xi_b,\oB)|0\rangle_A \otimes |0\rangle_B\,.
\end{align}
Clearly, the squeezing in both output arms of the beamsplitter will be lower than the one in the input mode.
As one can see in Eq. (\ref{eq:sqoutput}), the two output arms are squeezed coherent states.

The output state in the mode $\oA$ is written as 
\begin{align}\label{eq:dis_sq}
\ket{\psi_{out,A}} = D(k,\oA)S(T re^{i\omega}, A)|0\rangle_A\, ,
\end{align}
where $|k| = \alpha_b \sqrt{R'}$. It is noticeable that here amplitude and squeezing of the output field can be independently adjusted via the amplitude of the input fields, their relative phase and the beam splitter reflectivity. Accordingly to Ref. \cite{Lemonde2014}, this can allow one to minimize $g^{(2)}(0)$ function. 
It requires using an output state squeezed in the same direction as the displacement vector (squeezed in amplitude), and by choosing the amplitude of the output state as 
\begin{align}\label{eq:optcond}
|k| = \sqrt{\frac{\sinh(r/2)\sinh(r)}{e^{-3r/2}(e^{r}-1)}} . 
\end{align}

\begin{figure}[hbtp]
	\centering
	\includegraphics[width=0.9\linewidth]{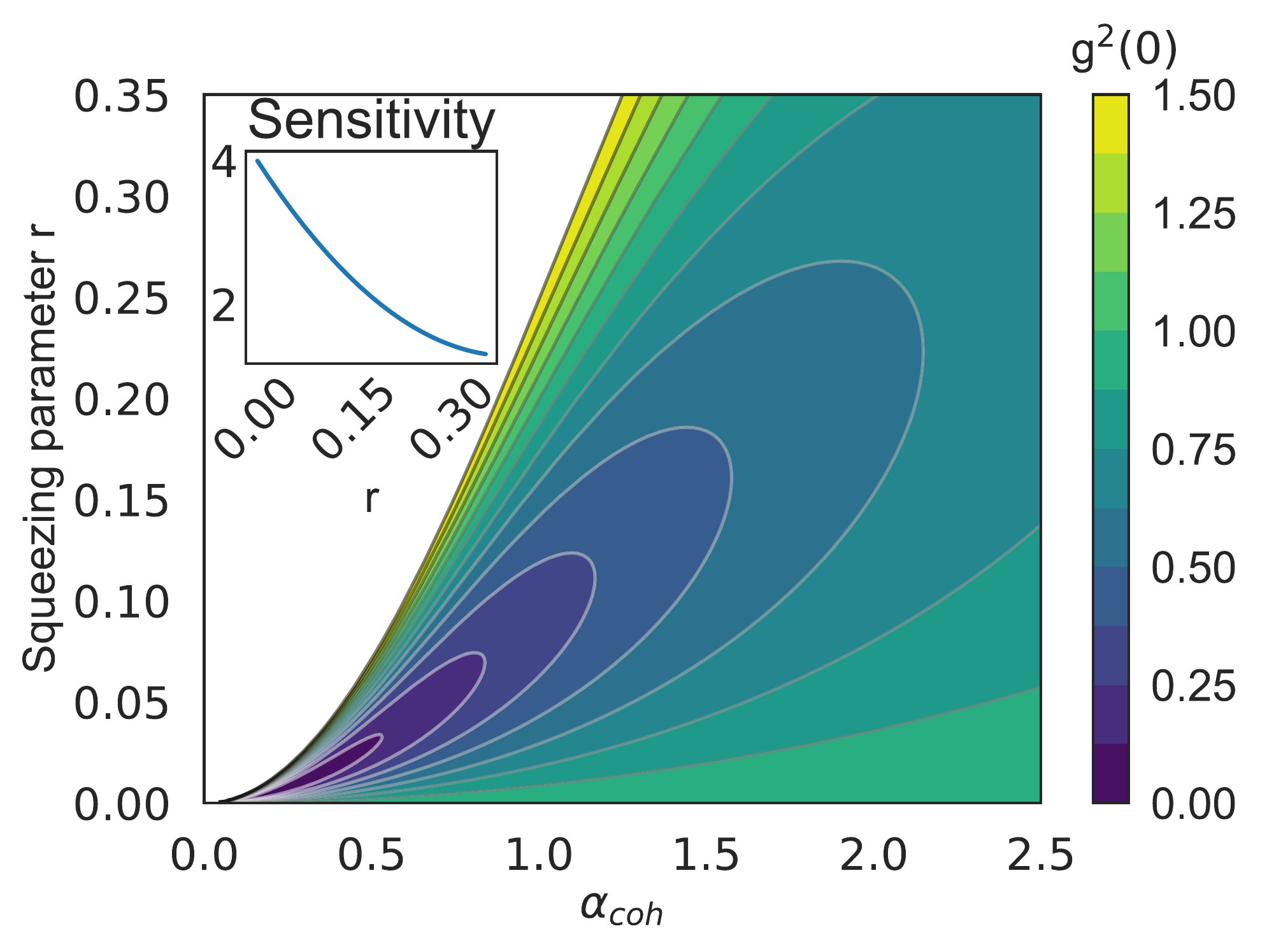}
	\caption{Contour plot of the second order correlation function in the output mode $\oA$ as a function of squeezing parameter $r$ of squeezed vacuum and the the amplitude of the coherent state in the input arms, $\alpha_{\text{coh}}$. We consider beamsplitter transmission $T = 90\%$ and $\phi = \pi$. It is also important to note that the squeezing direction is aligned with the coherent state. Inset: Sensitivity of the minima of the $g^{(2)}(0)$ as a function of the squeezing parameter$\sim r$. } \label{fig:g2ContourSqueezing}
\end{figure}


	
If there is only a squeezed vacuum state in one of the input arms ($\alpha_a=0$) and a coherent state in the other, the highest anti-bunching is found for 
\begin{eqnarray}
\phi&=& \arccos{\sqrt{T}}/2 - \Phi,\\
|\alpha_b| &=& \frac{1}{\sqrt{R}} e^{r'\sqrt{\frac{R}{T}}}\sqrt{\frac{\sinh(r')\sinh(2r')}{e^{-3r'}(e^{2r'}-1)}} ,
\end{eqnarray}
where $r'= r\sqrt{T}$ and $\alpha_b = |\alpha_b| e^{i\Phi}$.
The corresponding correlation function at zero delay $g^{(2)}(0)$ is shown in Fig. \ref{fig:g2ContourSqueezing}, as a function of the squeezing parameter and the amplitude of the input coherent state with a beamsplitter of 90\% transmission and a phase difference $\phi$ fixed to $\pi$. In the limit of intense coherent states, this experiment is analogous to the homodyne detection and $g^{(2)}(0)$ goes to 1.
However, in the limit of weak coherent states, as visible in Fig. \ref{fig:g2ContourSqueezing}, homodyne analogy is not valid anymore and interestingly the strongest anti-bunching is obtained for the weakest squeezing. 
As for the cat state case seen previously, there is a limit to the coherent state amplitude that allows for anti-bunching. The squeezing of the input field should be bigger than a critical value (increasing quickly with $\alpha$) to ensure anti-bunching in the output. More surprisingly, the correlation function vanishes when the squeezing parameter $r$ decreases. In analogy to what was shown in the previous section about cat states, mixing a coherent state with a state close to a coherent vacuum state leads to the strongest anti-bunching. Moreover, as for cat states, one can observe that the correlation function is more sensitive to fluctuations for lower squeezing (i.e. small $r$), and this property can be used to accurately measure weak squeezing.
 

Detecting squeezed states and cat states of light is a challenging task. The closer one gets to a coherent state, the setup's sensitivity should be higher and the more likely that one would run into technical difficulties \cite{Ourjoumtsev2011,glorieux2012generation}. The present setup goes actually the opposite way: getting closer to a coherent state the stronger is the signature on the $g^{(2)}(0)$ function. This setup can advantageously used to witness weakly squeezed or weak cat states. Moreover, scanning the amplitude of the input coherent field one can quantify with a very good accuracy via the $g^{(2)}(0)$ function how much squeezing is present in the input state or how big the input cat states are. The advantage of this method lies in its simplicity as it only requires a beamsplitter, a weak coherent beam and a coincidence measurement setup.

We will now show how this setup is related to unconventional photon blockade.

\section*{Unconventional photon blockade}

Here, we connect our results to the unconventional photon blockade. This phenomenon typically takes place in a coupled cavity system. It has been initially studied with both the cavities filled with $\chi^{(3)}$ non-linear medium and shown that the system outcomes will be a strong anti-bunched light for a specific set of parameter \cite{SavonaPRL}. Then it was understood that only one cavity needs to be filled with nonlinear medium to reach the same sub-poissonian statistics and that this strong anti-bunching results from interfering optical paths \cite{Flayac2017, Bamba2011}.
We have presented in the section related to phase modified coherent states, that using just one cavity filled with non-linear medium can lead to the same strong anti-bunching. 

\begin{figure}[hbtp]
	\centering 
	\includegraphics[width=\linewidth]{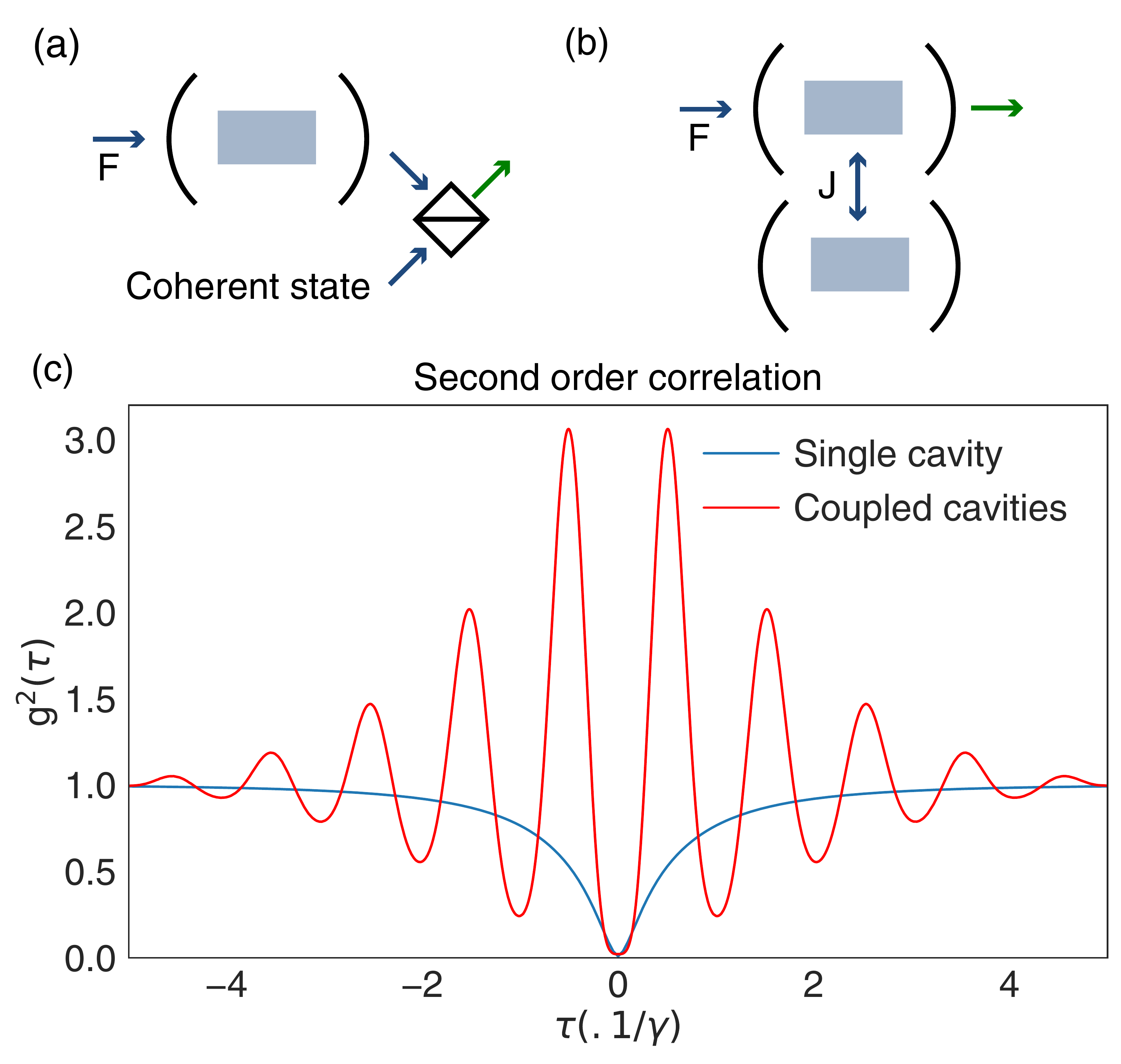}
	\caption{Here we compare two systems one consists of a cavity with a non linear medium of non-linearity U = 0.01$\gamma$ and the other with coupled cavity filled with same non-linear medium as illustrated in figure (a) and (b). $\gamma$ is the cavity linewidth, $J=6.2\gamma$ is the coupling parameter, F is the feeding rate into the system and we evaluate $g^{(2)}(\tau)$ in the output path denoted in green, normalized to the cavity lifetime. All the other parameters are adjusted to have the best anti-bunching in the system.} \label{fig:SqCav}
\end{figure}

In order to compare our results with the original proposal with two coupled cavities \cite{SavonaPRL,Bamba2011}, we modify the setup proposed adding a cavity of linewidth $\gamma$, filled with a non-linear medium with a non-linearity of 0.01$\gamma$ before the beam splitter (c.f. Fig. \ref{fig:SqCav}.a). For these parameters, we have numerically estimated the amount of squeezing after the cavity to  about 1\%. As discussed in previous section this allows for a very strong anti-bunching after mixing adequately with a coherent state.

In Fig \ref{fig:SqCav}.c, the correlation function at the output of the setup is plotted in blue as a function of the time delay, normalized to the cavity lifetime. We observe strong anti-bunching dip, with a linewidth of the same order as the cavity linewidth. 

For comparison, we plot in red the correlation function one would expect for the same non-linearity in the coupled cavity case (Schematic of the coupled cavity setup in Fig. \ref{fig:SqCav}.b). One can clearly observe the contrasting behaviour of $g^{(2)}(\tau)$. If, as for a single cavity setup the $g^{(2)}(\tau)$ function almost vanished for $\tau \to 0$, for finite delay it strongly differ. Whereas single cavity leads to a monotonous increase of the $g^{(2)}(\tau)$ to reach $1$, the two cavity setup strongly oscillate at a frequency equal to the cavity linewidth. Moreover, the amplitude of the oscillation greatly overcomes the shot noise value and vanishes for long time delay.

Strictly speaking anti-bunching can only be defined for a time window in which $g^{(2)}(\tau)$ increases monotonously with $|\tau|$ and not only for $g^{(2)}(0)<1$ \cite{Zou1990}. Consequently, 
as it can be seen in Fig. \ref{fig:SqCav}.c in the coupled cavities case (red), one would require to consider only the time window such as  $|\tau|\le \gamma/2$ to actually have sub-poissonian statistics. Hence, in such configuration a complex time filtering is necessary to create strongly anti-bunched state of light.
As already pointed out by H. Flayac and coworkers in \cite{Flayac2017}, one can overcome this important limitation and obtain similarly strong anti-bunching with a single cavity mixing its output with a coherent state on a beamsplitter as clearly illustrated in Fig. \ref{fig:SqCav}.c.

\section*{Conclusion}
In this article, we focus on a key measurement in quantum optics commonly used to characterize single photon source, the second order correlation function, $g^{(2)}(0)$. We showed that using a rather simple setup one can generate strongly anti-bunched states of light ($g^{(2)}(0)\ll 1$). This setup based on simple beamsplitter when it is used to mix a coherent field with $g^{(2)}(0)=1$ with another state characterized by $g^{(2)}(0)>1$, can provide an output field such as $g^{(2)}(0)<1$. We reveal how this mechanism is a consequence of interfering different Fock states components of the input beams. We considered experimentally feasible conditions and detailed how this setup can be advantageously applied to characterize weak squeezed states and Schr\"{o}dinger cat states. Finally, we connected our results to the unconventional photon blockade to show that both phenomena rely on the same physics. 
This work is important as it offers a simple setup to generate on demand anti-bunched states of light, which had been found to have many rich applications in the last decade \cite{OBrien2009}. Moreover, thanks to its simplicity, the proposed scheme can be easily and efficiently integrated to become a central piece of emerging quantum technologies.  

\section*{Acknowledgements}
The authors would like to thank Cristiano Ciuti, Alejandro Giacomotti for fruitful discussions and Elisabeth Giacobino for her comments on the manuscript. This work is supported by the French ANR grants (UNIQ DS078, C-FLigHT 138678).
\bibliographystyle{apsrev4-1}
\bibliography{Bibliography}
\end{document}